\documentclass[conference]{IEEEtran}
\IEEEoverridecommandlockouts
\usepackage{cite}
\usepackage{amsmath,amssymb,amsfonts}
\usepackage{algorithmic}
\usepackage{graphicx}
\usepackage{textcomp}
\usepackage{xcolor}
\usepackage{soul}
\usepackage{array}
\usepackage{tabularx}
\usepackage{float}
\usepackage{tcolorbox} 
\usepackage{listings}  

\def\BibTeX{{\rm B\kern-.05em{\sc i\kern-.025em b}\kern-.08em
    T\kern-.1667em\lower.7ex\hbox{E}\kern-.125emX}}
\begin{document}

\title{Zero-Shot Cognitive Impairment Detection from Speech Using AudioLLM}

\author{\IEEEauthorblockN{Mostafa Shahin}
\IEEEauthorblockA{\textit{University of New South Wales}\\
m.shahin@unsw.edu.au}
\and
\IEEEauthorblockN{Beena Ahmed}
\IEEEauthorblockA{\textit{University of New South Wales}\\
beena.ahmed@unsw.edu.au}
\and
\IEEEauthorblockN{Julien Epps}
\IEEEauthorblockA{\textit{University of New South Wales}\\
j.epps@unsw.edu.au}
}

\maketitle

\begin{abstract}
Cognitive impairment (CI) is of growing public health concern, and early detection is vital for effective intervention. Speech has gained attention as a non-invasive and easily collectible biomarker for assessing cognitive decline. Traditional CI detection methods typically rely on supervised models trained on acoustic and linguistic features extracted from speech, which often require manual annotation and may not generalize well across datasets and languages. In this work, we propose the first zero-shot speech-based CI detection method using the Qwen2-Audio AudioLLM—a model capable of processing both audio and text inputs. By designing prompt-based instructions, we guide the model to classify speech samples as indicative of normal cognition or cognitive impairment. We evaluate our approach on two datasets: one in English and another multilingual, spanning different cognitive assessment tasks. Our results show that the zero-shot AudioLLM approach achieves performance comparable to supervised methods, and exhibits promising generalizability and consistency across languages, tasks, and datasets.
\end{abstract}

\begin{IEEEkeywords}
AudioLLM, Cognitive Impairment, Zero-Shot
\end{IEEEkeywords}

\section{Introduction}
Early detection of cognitive impairment (CI) is crucial, as it often serves as an early indicator of Alzheimer’s disease and other conditions associated with cognitive decline \cite{AssociationAlzheimers20232023Figures}. Identifying CI at an early stage enables timely interventions that can slow its progression, enhance quality of life, and improve patient management.
Recognizing early signs of CI involves detecting subtle changes in memory, language, and decision-making, such as frequent forgetfulness, difficulty retrieving words, or impaired judgment \cite{Nowrangi2016SubtleDatabase}. However, for elderly individuals — especially those living alone — these signs may go unnoticed, delaying diagnosis and intervention \cite{Gamble2022CharacteristicsStudy}.

Speech and language are among the earliest cognitive abilities affected by CI, often manifesting as disfluencies, word-finding difficulties, frequent pauses, and a reduced vocabulary. Since speech impairments occur at various stages of the disease, they play a critical role in distinguishing disease progression and forecasting its trajectory. Moreover, speech is a non-invasive, easily collectible biomarker that allows for continuous and remote monitoring, reducing the need for costly, time-consuming and typically infrequent in-person clinical assessments \cite{Ding2024SpeechChallenges}. Automated speech based CI detection can significantly streamline the diagnosis process and enable early intervention \cite{DeLaFuenteGarcia2020ArtificialReview}.

Current methods for automatic CI detection from speech typically rely on two sets of features: acoustic features, such as low-level descriptors (LLDs) \cite{Martinc2020TacklingDementia, Luz2020AlzheimersChallenge, Edwards2020MultiscaleSpeech, Luz2024ConnectedEnglish}, which capture paralinguistic characteristics of the speech signal; and linguistic features, mainly text embeddings \cite{Edwards2020MultiscaleSpeech, Martinc2020TacklingDementia, Ortiz-Perez2024CognitiveAnalysis}, which analyze the spoken content. Linguistic features are extracted either from manually annotated transcripts (when available) \cite{Luz2020AlzheimersChallenge} or from transcripts generated by automatic speech recognition (ASR) systems \cite{Luz2024ConnectedEnglish, AndreaPerez-Toro2024MultilingualChallenge, Ortiz-Perez2024CognitiveAnalysis}.

Despite advances in dataset availability, CI detection remains constrained by data scarcity. Although several datasets for cognitive assessment have been released, such as ADReSS \cite{Luz2020AlzheimersChallenge}, Pitt \cite{Becker1994TheDiagnosis}, TAUKADIAL\cite{Luz2024ConnectedEnglish}, and PROCESS \cite{Tao2025EarlyChallenge}, the total amount of data remains limited. One of the primary barriers to dataset expansion is privacy concern, as cognitive assessment data is inherently sensitive and challenging to share. Furthermore, most existing models are trained and evaluated on in-domain datasets, making generalization across different datasets and populations difficult. This highlights the need for robust models capable of handling diverse and unseen data, such as zero-shot and few-shot learning approaches.

To address these challenges, we propose a zero-shot CI detection approach using Qwen2-Audio, an Audio Large Language Model (AudioLLM) capable of processing both speech and text. AudioLLMs build on the success of Large Language Models (LLMs), which have revolutionized natural language processing by enabling advanced text understanding and generation. AudioLLMs \cite{Peng2024AMODELS}, integrate speech processing capabilities into LLMs, allowing them to analyze and interpret both text and audio inputs. These models are built on top of existing LLMs, with WavLLM \cite{Hu2024WavLLM:Model} and SALMONN \cite{Tang2024SALMONN:MODELS} based on Llama models, while Qwen-Audio \cite{Chu2023Qwen-Audio:Models} and Qwen2-Audio \cite{Chu2024Qwen2-AudioReport} are built on Qwen LLM \cite{Bai2023QwenReport}.

 We designed various prompt types to instruct the model to classify speech signals as either normal cognitive decline (NC) or CI. To evaluate the effectiveness of this approach, we conducted experiments on two datasets covering both English and Mandarin languages and incorporating different cognitive assessment tasks. The key contributions of this paper are as follows:

\begin{itemize}
    
\item Proposed the first zero-shot, speech-based method for cognitive impairment detection, achieving performance comparable to supervised approaches.
\item Demonstrated the ability of AudioLLM to follow natural language prompts for analyzing speech and detecting cognitive impairment.
\item Evaluated the effect of different prompt types, showing that prompts with richer contextual information improve detection accuracy.
\item Provided evidence of the model’s generalizability and consistency across different languages, cognitive tasks, and datasets.
\end{itemize}
\section{Background}
\subsection{Speech-based CI detection}

Speech is considered a convenient and non-invasive biomarker for cognitive decline, making it an attractive modality for automated CI detection. As a result, numerous studies have explored the use of speech analysis for this purpose. Current approaches typically rely on two categories of features: acoustic features, which capture paralinguistic aspects of the speech signal such as prosody and pauses, and linguistic features, which reflect the content and structure of the spoken language. Linguistic features are either derived from manually annotated transcripts when available \cite{Luz2020AlzheimersChallenge}, or extracted using ASR systems \cite{Luz2024ConnectedEnglish,AndreaPerez-Toro2024MultilingualChallenge,Ortiz-Perez2024CognitiveAnalysis}.

The most commonly used acoustic features include LLDs, which measure paralinguistic attributes such as pitch variability, speech rate, voice quality, and prosodic rhythm. These features are extracted at the frame level and are typically aggregated using statistical functionals such as mean, standard deviation, and percentiles. Standard acoustic feature sets include eGeMAPs \cite{Luz2024ConnectedEnglish,Martinc2020TacklingDementia,Edwards2020MultiscaleSpeech}, ComParE \cite{Cummins2020ARecognition,Zafar2025Multi-ClassChallenge}, and MRCG \cite{Luz2020AlzheimersChallenge, Edwards2020MultiscaleSpeech}.

One of the most effective acoustic feature categories is speech timing-related features, which capture pause distributions and speech intervals \cite{AndreaPerez-Toro2024MultilingualChallenge,Yuan2020DisfluenciesDisease}. Although these are contained in some LLDs, some studies have focused exclusively on timing-related features for CI detection \cite{AndreaPerez-Toro2024MultilingualChallenge}, as prolonged pauses and hesitations are strong indicators of cognitive decline.

With the rise of speech foundation models, recent research has leveraged deep learning-based speech embeddings for feature extraction. Commonly used foundation models in CI diagnosis include Wav2Vec2 \cite{Baevski2020Wav2vecRepresentations}, HuBERT \cite{Hsu2021HuBERT:Units}, and the Whisper encoder \cite{Radford2022RobustSupervision}, which generate rich representations of speech signals for downstream classification tasks \cite{AndreaPerez-Toro2024MultilingualChallenge,Favaro2024LeveragingLanguages, Gosztolya2024CombiningChallenge, Kurtz2023EarlyCommands}.

Linguistic features are typically extracted by generating text embeddings, using either traditional methods \cite{Martinc2020TacklingDementia,Edwards2020MultiscaleSpeech} such as word2vec \cite{Bojanowski2017EnrichingInformation}, GloVe \cite{Pennington2014GloVe:Representation}, and sent2vec \cite{Gupta2019BetterInformation}, or more recent transformer-based approaches, primarily BERT-based embeddings \cite{Yuan2020DisfluenciesDisease,Barrera-Altuna2024TheApproach,Hoang2024TranslingualSpeech, Balagopalan2020ToDetection}. 

Both acoustic and linguistic features are combined and fed into classification models, which are often conventional machine learning models such as Support Vector Machines (SVM), Logistic Regression , Random Forest, and Multi-Layer Perceptrons (MLP) \cite{Martinc2020TacklingDementia, Balagopalan2020ToDetection}. Some studies have explored deep learning architectures, including Long Short-Term Memory (LSTM) networks \cite{Luz2020AlzheimersChallenge, Rohanian2020Multi-modalSpeech}, Convolutional Neural Networks (CNNs) \cite{Cummins2020ARecognition}, and attention-based models \cite{Duan2024Pre-trainedDetection}.

Speech data for CI detection is typically collected by instructing individuals to perform specific speech tasks. One of the most commonly used tasks is picture description, where participants are asked to describe the content of an image or narrate a coherent story based on it. The "Cookie Theft" picture, introduced in the Boston Diagnostic Aphasia Examination \cite{Goodglass1983BostonBooklet}, is the most frequently employed image for this purpose. Other widely used tasks include word fluency \cite{Tao2025EarlyChallenge,Becker1994TheDiagnosis} and memory recall \cite{Becker1994TheDiagnosis}, which elicit speech samples that are informative for evaluating cognitive function. However, the number of publicly available datasets for CI speech analysis remains limited, primarily due to privacy concerns that restrict many medical institutions from sharing their data. These limitations hinder the development of robust speech-based CI detection systems that can generalize across diverse populations, tasks, and recording conditions.

\subsection{AudioLLM}

In 2017, the Google Research team introduced the Transformer architecture in their seminal paper, "Attention Is All You Need."\cite{Vaswani2017AttentionNeed}. This model revolutionized deep learning by replacing traditional recurrent and convolutional architectures with an attention mechanism \cite{Bahdanau2014NeuralTranslate}. The encoder-decoder structure of Transformers enables efficient parallel processing, significantly improving computational efficiency and scalability.

Since then, the Transformer architecture has become the foundation of modern artificial intelligence, particularly in natural language processing (NLP). Its ability to process vast amounts of text data has enabled the development of Large Language Models (LLMs) such as BERT\cite{Devlin2018BERT:Understanding}, GPT\cite{Openai2018ImprovingPre-Training}, and QwenLLM\cite{Bai2023QwenReport}, which power state-of-the-art applications in machine translation, summarization, text generation, and speech understanding.

Current LLMs are trained on text data from diverse domains and disciplines, including medical data that contains information on various medical conditions, diseases, diagnoses, treatments, and clinical guidelines. This broad knowledge has led to research exploring their potential in healthcare applications, including assisting in diagnosis and medical research \cite{Thirunavukarasu2023LargeMedicine, Savage2024DiagnosticMedicine}.

Following the success of text-based LLMs in NLP tasks, research has expanded toward multimodal LLMs, which can process and generate outputs in multiple formats, including text, images, video, and audio.
A key category within multimodal LLMs is AudioLLMs \cite{Peng2024AMODELS}, which are designed to process both audio and text inputs. These models extend a pretrained LLM by incorporating one or more audio encoders to generate representations of the audio signal before passing it to the LLM for interpretation and response generation.

In WavLLM \cite{Hu2024WavLLM:Model}, the audio input is processed through two audio encoders—wavLM \cite{Chen2021WavLM:Processing} and Whisper \cite{Radford2022RobustSupervision}. wavLM captures speaker-related acoustic features, while Whisper encodes semantic information. The output from each encoder is passed through a dedicated trainable adapter—an acoustic adapter for wavLM and a semantic adapter for Whisper. The outputs from these adapters are then concatenated and passed through a linear layer before being fed into the Llama3-based LLM backbone. Additionally, WavLLM employs a prompt-aware adaptive LoRA, which dynamically adjusts the LLM’s responses based on the task-specific prompt using the LoRA \cite{Hu2021LORA:MODELS} fine-tuning technique.

Similarly, SALMONN AudioLLM \cite{Tang2024SALMONN:MODELS} follows a dual audio encoder + LLM architecture. It utilizes Whisper \cite{Radford2022RobustSupervision} to capture speech-related semantic information and BEATs \cite{Chen2022BEATs:Tokenizers} to extract non-speech audio semantics. The outputs of these two encoders are then fused through a stacked series of Q-Former blocks, which convert them into a fixed set of textual tokens before passing them to the LLM. SALMONN employs Vicuna \cite{Zheng2023JudgingArena}—a fine-tuned LLaMA model designed for instruction-following tasks—and further adapts it using LoRA-based parameter-efficient fine-tuning.

By contrast, Qwen-Audio \cite{Chu2023Qwen-Audio:Models} follows a simpler architecture. It builds upon Qwen LLM and uses a single Whisper audio encoder to encode the input audio. The encoded representations are then concatenated with the tokenized prompt before being passed to the LLM for processing.

As AudioLLMs are a relatively new advancement, limited efforts have been made to systematically evaluate their effectiveness across various tasks \cite{Wang2024AudioBench:Models,Chen2024VoiceBench:Assistants}. One such effort is AudioBench \cite{Wang2024AudioBench:Models}, which introduces a comprehensive benchmark consisting of diverse datasets and evaluation procedures. AudioBench assesses these models by prompting them either through text-based instructions or by appending audio-based instructions to the input audio file. The benchmark evaluates models across three key task categories: speech understanding, which includes tasks like speech-to-text and speech question answering; audio scene understanding, covering tasks such as audio captioning and audio-scene question answering; and voice understanding, which focuses on attributes like accent, gender, and emotion recognition. 

However, these models have not yet been tested on medical-related speech tasks, including the detection of cognitive behavior. In this work, we present the first attempt to evaluate the Qwen2-Audio AudioLLM for detecting cognitive status from speech, where the model processes speech signals and is guided by text-based prompts.

\section{Method}
Our method leverages AudioLLMs capable of processing both audio and text inputs to perform zero-shot classification of cognitive status, categorizing speech samples as either NC or CI. The audio input consists of speech recordings from elderly individuals performing various cognitive assessment tasks, while the text input serves as a prompt instructing the AudioLLM to analyze and classify the speech accordingly.

For this study, we employed Qwen2-Audio \cite{Chu2024Qwen2-AudioReport}, which has demonstrated superior performance compared to other AudioLLMs in several speech and voice understanding tasks, including emotion and sentiment recognition \cite{Wang2024AudioBench:Models}. These tasks require the analysis of paralinguistic elements—such as prosody, tone, and hesitations—aligning closely with the characteristics relevant to cognitive assessment.

\subsection{Datasets}
This work utilizes two datasets: the TAUKADIAL dataset \cite{Luz2024ConnectedEnglish} and the PROCESS dataset \cite{Tao2025EarlyChallenge}. Both datasets contain recordings from cognitive assessment sessions in which a clinician instructs participants to perform specific speech tasks.

In the TAUKADIAL dataset, participants were asked to describe a picture. The dataset includes speech from male and female speakers aged 61–87 years, speaking either English or Mandarin. The participants are categorized into two groups: those with normal cognitive function (NC) and those diagnosed with mild cognitive impairment (MCI). This dataset was introduced as part of a challenge at Interspeech 2024 \cite{Luz2024ConnectedEnglish}.

The dataset includes an official train-test split. However, since our approach relies on zero-shot classification using an AudioLLM, we merge the train and test sets for evaluation. When comparing our results with the official baseline, we computed metrics separately on the test set to enable a direct comparison with the baseline model performance reported in the dataset’s official release paper.
The dataset consists of 507 speech samples, with 285 from MCI participants and 222 from NC participants.

In contrast, the PROCESS dataset includes three different tasks:
\begin{itemize}
    \item Cookie Theft Picture Description (CTD): Participants describe a well-known image used in cognitive assessments.
    \item Semantic Fluency Task (SFT): Participants name as many animals as possible within one minute.
    \item Phonemic Fluency Task (PFT): Participants generate as many words as possible beginning with the letter P within one minute.
\end{itemize}

Unlike TAUKADIAL, the PROCESS dataset contains three groups of participants: NC, MCI, and individuals with dementia. To unify the classification labels across both datasets, we merged the dementia and MCI groups into a single CI category, resulting in a binary classification task (NC vs. CI). This allows the use of a consistent set of prompts for both datasets.
The PROCESS dataset was introduced as a challenge in ICASSP 2025 \cite{Tao2025EarlyChallenge}. The training and development sets include 157 participants: 82 NC, 59 MCI, and 20 with dementia. Similarly to the TAUKADIAL dataset, we merged the training and testing subsets for evaluation. 

Table \ref{tab:datasets} provides a detailed breakdown of both datasets.
In both datasets, the metadata includes information about the age and gender of each speaker, which we incorporated into some prompts to provide the AudioLLM model with additional contextual information.

\begin{table}[htbp]
    \caption{Statistics of the two datasets used in this study. TAUKADIAL includes two languages (English and Mandarin) and three Picture Description (PD) tasks. PROCESS is an English-only dataset with three different tasks: Picture Description (PD), Semantic Fluency (SFT), and Phonetic Fluency (PFT).}
    \begin{center}
    \begin{tabular}{|c|c|c|c|}
    \hline
         \textbf{Dataset} & \textbf{Tasks} & \textbf{Languge(s)} & \textbf{N\# Subjects} \\
         & & & Healthy/CI \\
         \hline
         TAUKADIAL & PD & English, Mandarin & 165/222 \\
         \hline
         PROCESS & PD, SFT, PFT & English & 82/79 \\
         \hline
    \end{tabular}
    \label{tab:datasets}
    \end{center}
\end{table}

These two datasets were selected as they are among the most recent benchmark datasets, each introduced as part of a challenge with clearly defined baselines, enabling a direct comparison of our proposed method. Additionally, TAUKADIAL includes two languages, allowing us to evaluate performance across different linguistic contexts. Meanwhile, PROCESS features multiple tasks, providing an opportunity to assess how the model performs across different cognitive assessment tasks.

\subsection{Qwen2-Audio}
\begin{figure}[htbp]
\centering 
\includegraphics[width=1\linewidth]{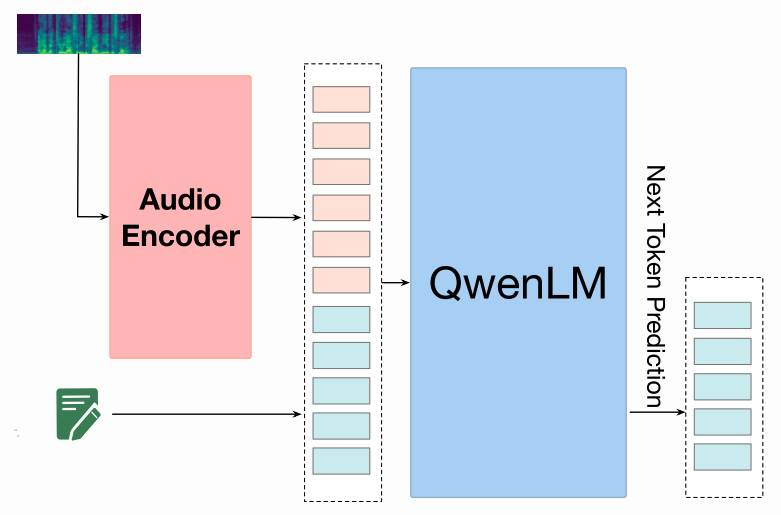}
\caption{The architecture of Qwen2-Audio. The model is built on the QwenLLM. The input audio is first encoded using an audio encoder, typically the Whisper-large-v3 encoder, before being concatenated to prompt tokens and fed into the QwenLLM for processing.}
\label{fig:qwen2}
\end{figure}

In this work, we adopted the Qwen2-Audio model \cite{Chu2024Qwen2-AudioReport}, the second version of the Qwen AudioLLM series  \cite{Chu2023Qwen-Audio:Models}. Qwen AudioLLM is a speech-and-text-to-text (ST2T) model that takes both an audio signal and a text prompt as input and generates text as output. The model architecture, illustrated in Fig. \ref{fig:qwen2}, consists of an audio encoder that processes features extracted from the input audio signal and generates a set of output representations. These representations were then concatenated to the tokenized text prompt and fed into the Qwen LLM, which serves as the backbone of the model.

Qwen2-Audio utilizes the Whisper-large-v3 encoder for audio encoding and the Qwen-7B LLM \cite{Qwen2024Qwen2.5Report} for text generation. The model was trained on approximately 30 different audio, speech, and music tasks, including ASR, emotion recognition (ER), age prediction (SAP), speaker recognition (SR), speaker diarization (SD), and more.

The training process consists of multiple stages. The first stage, multi-task pretraining, involves training the model on various tasks by providing different audio-text input pairs. For example, an audio input may be paired with a text prompt requesting transcription (ASR task) or with a prompt asking to detect speaker emotion (ER task). During this stage, the LLM parameters remain frozen, and only the audio encoder parameters are updated.

In the supervised fine-tuning stage, the model is trained in two operational modes. In audio analysis mode, it analyzes input audio and responds to queries based on given instructions, which may be provided as a text prompt or embedded within the input audio. In chat mode, the model is trained to engage in interactive conversations with users.

The final stage, direct preference optimization (DPO) fine-tunes the model to align with human preferences. This involves a supervised training process using triplet samples — an input, a good response, and a bad response — where the model is optimized to assign a higher score to the preferred response and a lower score to the undesirable one.
 
\subsection{Prompt Design}
\label{sec:prompt}
Since the goal of this work was to explore the ability of Qwen2-Audio to assess cognitive behavior from speech — without explicit fine-tuning for this task — we designed five types of prompts, as shown in Table \ref{tab:prompts}.

\begin{table*}[htbp]
    \caption{Five types of prompts: three direct prompts with varying levels of contextual information and two Chain of Thought (CoT) prompts.}
    \begin{center}
    \begin{tabularx}{\textwidth}{|c|X|}
    \hline
    \textbf{Prompt Type} & \textbf{Example} \\
    \hline
    \textbf{Direct} & From the given speech sample, determine the cognitive condition. Select one of the following labels: NC for Normal Cognitive Decline or MCI for Mild Cognitive Impairment. Reply with only one word: NC or MCI \\
    \hline
    \textbf{Contextual} & Assess the cognitive condition based on the input audio, where an elderly speaker describes one of three images as part of a clinician-guided task. Indicate the diagnosis using one of these labels: NC for Normal Cognitive Decline or MCI for Mild Cognitive Impairment. Output only NC or MCI as your response. \\
    \hline
    \textbf{Informaive} & Assess the cognitive condition based on the input audio, in which a [AGE]-year-old [GENDER] describes a picture in [LANGUAGE] as part of a clinician-guided task. Indicate the diagnosis using one of these labels: NC for Normal Cognitive Decline or MCI for Mild Cognitive Impairment. Output only NC or MCI as your response. \\
    \hline
    \textbf{CoT} & First, convert the speech in the input audio into text internally, but do not output the transcription. Then, analyze both the acoustic and linguistic features to determine the cognitive diagnosis. Provide only one word as the answer: NC or MCI. \\
    \hline
    \textbf{CoT-Informative} & First, transcribe the speech from the input audio internally. The audio is part of a cognitive test, where elderly individuals describe images as prompted by clinicians. Then, analyze both the audio and the transcription to determine the diagnosis. Do not output the transcription. Respond with only one word: NC for Normal Cognitive Decline or MCI for Mild Cognitive Impairment. Output only NC or MCI. \\
    \hline
    \end{tabularx}
    \label{tab:prompts}
    \end{center}
\end{table*}

The first type (Direct) is a straightforward classification prompt. The prompt provides minimal context about the task or the content of the target audio file and directly asks the model to classify the speaker's cognitive ability. The objective is to encourage the model to rely primarily on acoustic cues in the audio.

The second type (Contextual) introduces additional context by specifying that the speech file is part of a cognitive assessment and that the speaker is an elderly individual. The prompt also includes information about the task being performed. In the TAUKADIAL dataset, where the only task is image description, the prompt explicitly states that the speaker is describing an image. In the PROCESS dataset, which includes three different tasks — describing the Cookie Theft image, naming animals, and generating words that start with the letter P — each task has a specific prompt that clarifies the nature of the task. This prompt type allows us to examine whether such contextual information helps the model better understand the classification task.

The third type (Informative) builds on the contextual prompt by incorporating demographic details such as age, gender, and language. In the TAUKADIAL dataset, the language can be either English or Mandarin, whereas in the PROCESS dataset, it is always English. By providing these details, we investigate whether demographic information enhances the model’s classification performance.

In the first three prompt types described above, the model is asked to classify the audio file based on the speaker’s cognitive ability without being guided through a structured reasoning process. In contrast, the last two prompt types use a Chain of Thought (CoT) approach, requiring the model to perform intermediate steps before reaching a final decision. Since cognitive impairment can be assessed through both speech cues (e.g., pauses, disfluencies) and linguistic features (e.g., vocabulary, word selection), these prompts first instruct the model to generate a text transcription of the speech. The model is then asked to analyze both the audio and text before making an assessment. This approach explicitly guides the model to consider both acoustic and linguistic features.

The difference between the two CoT prompts is that one includes only the transcription and analysis steps, while the other incorporates additional contextual information about the task, similar to the Contextual and Informative prompt types. As these prompts contain multiple instructions, the model sometimes struggles to output a single-word response. To ensure that the classification remains binary (i.e., NC for cognitively normal or MCI for mild cognitive impairment), we explicitly emphasized that all intermediate steps should be performed internally, with the model responding only with a single-word classification.

To assess the impact of rephrasing on the model's response, we also asked the ChatGPT-4.0 model to generate five variants of each prompt, preserving the same core concept of each type. This enabled us to investigate how different phrasings might influence the model's classification performance and consistency.

\section{Experimental Setup}
We used the \emph{Qwen2-Audio-7B-Instruct} model from the Hugging Face repository\footnote{https://huggingface.co/Qwen/Qwen2-Audio-7B-Instruct}, loading it with the \emph{Qwen2AudioForConditionalGeneration} transformer class. The model has approximately 8.2B parameters, with 7B allocated to the QwenLLM and the remainder for the Whisper audio encoder. The full-precision (fp32) version of the model requires around 32 GB, while the half-precision (fp16) version reduces this to 15 GB. For inference, we ran the model on an NVIDIA V100 GPU with 32 GB of memory, loading it in fp16 mode.
All audio files were converted to single-channel, 16 kHz format to ensure compatibility with Qwen2-Audio.

Table \ref{tab:prompt_template} presents the template prompt used during inference. The prompt follows a conversational format, with roles enclosed between $|im\_start|$ and $|im\_end|$. The system role instructs the model to act as an assistant, followed by the user role, which consists of both text and audio components. The audio input is enclosed between the special tokens $<|audio\_bos|>$ and $<|audio\_eos|>$, marking the beginning and end of the audio stream. The final assistant role ($|im\_start|$ assistant) acts as a generation prompt, ensuring that the model produces a response immediately afterward.

\begin{table}[H]
    \centering
    \caption{Structured Qwen2-Audio prompt template}
    \label{tab:prompt_template}

\begin{tcolorbox}[colback=gray!10, colframe=black, title=Prompt Template]
\begin{verbatim}
<|im_start|>system
You are a helpful assistant.
<|im_end|>
<|im_start|>user
Audio 1: <|audio_bos|>
<|AUDIO|>
<|audio_eos|>
[PROMPT]
<|im_end|>
<|im_start|>assistant
\end{verbatim}
\end{tcolorbox}
\end{table}

\section{Results}
As shown in Table \ref{tab:prompts}, we designed five prompt types: three direct prompts with varying levels of contextual information (Direct, Contextual, Informative) and two Chain of Thought (CoT) prompts, with (CoT-Informative) and without (CoT) contextual information. For each prompt type, five reworded variants were generated while maintaining the same prompting strategy.

Table \ref{tab:results_tkd} presents the Unweighted Average Recall (UAR) and macro-F1 scores for the TAUKADIAL dataset across all prompt types, reporting only the best-performing prompt (based on UAR) within each type. Among the direct prompts, adding contextual information generally improved performance, with Contextual and Informative prompts outperforming the Direct prompt. Similarly, CoT-Informative performed better than CoT without context. However, CoT prompting did not offer an advantage over direct prompting in this setup. The best performance was achieved using the Contextual prompt, with 57.5\% UAR and 56.8\% macro-F1.

\begin{table}[htbp]
    \caption{Unweighted Average Recall (UAR) and macro-F1 (mF1), as \%, of Qwen2-Audio for binary classification of the TAUKADIAL dataset into NC/MCI using different prompt types. The best performance was achieved with the Contextual prompt type. Overall, the Mandarin portion of the dataset outperforms the English portion.}
    \label{tab:results_tkd}
    \begin{center}
        \begin{tabular}{|c|c|c|c|c|c|c|}
        \hline
         \textbf{Prompt Type} &  \multicolumn{2}{|c|}{\textbf{Overall}} & \multicolumn{2}{|c|}{\textbf{English}} & \multicolumn{2}{|c|}{\textbf{Mandarin}}\\
         \cline{2-7}
         & \textbf{UAR} & \textbf{mF1} & \textbf{UAR} & \textbf{mF1} & \textbf{UAR} & \textbf{mF1} \\
         \hline
         \textbf{Direct} & 52.17 & 43.38 & 50.07 & 42.28 & 53.91 & 43.66 \\
         \hline
         \textbf{Contextual} & \textbf{57.5} & \textbf{56.79} & 54.6 & 52.34 & 60.94 & 60.91 \\
         \hline
         \textbf{Informative} & 53.21 & 51.32 & 50.48 & 49.61 & 56.01 & 52.61 \\
         \hline
         \textbf{CoT} & 51.55 &	44.21 &	51.88 &	47.35 &	51.62 &	41.30 \\
         \hline
         \textbf{CoT-Informative} & 53.30 &	49.65 &	51.95 &	41.49 &	57.20 &	56.75 \\
         \hline
    \end{tabular}
    
    \end{center}
    
\end{table}

We further break down the evaluation by language in the TAUKADIAL dataset, comparing performance on English and Chinese samples. As shown in Table \ref{tab:results_tkd}, the model generally performs better on Chinese data across almost all prompt types, with a more significant improvement observed in contextual prompts. One possible explanation for this trend is that Qwen2-Audio was trained on a substantial amount of Mandarin data, as reported in \cite{Chu2024Qwen2-AudioReport}.

For comparison with the official baseline models reported in \cite{Luz2024ConnectedEnglish}, we evaluated the model on the test set separately for each prompt type. To further enhance classification, we applied Majority Vote (MV) across the five best-performing prompts (one per type). Using the training split, we selected the best prompt within each type, applied the same selection to the test set, and aggregated predictions using the MV rule. This improved the test UAR from 55.2\% (best individual Contextual prompt) to 59

Table \ref{tab:results_tkd_test} presents the Qwen2-Audio LLM performance alongside the baseline performance of the TAUKADIAL test split. The baseline model is a binary MLP classifier trained on hand-crafted speech features. Notably, Qwen2-Audio’s MV results outperformed nearly all baseline models, despite being a zero-shot system, whereas the baseline models were trained on in-domain data.

\begin{table*}[htbp]
 \caption{Comparison of UAR(\%) on the TAUKADIAL test set between the Qwen2-Audio model (using different prompt types) and various baseline models \cite{Luz2024ConnectedEnglish}. In the baseline models, "w2v" refers to Wav2Vec features, "ling" represents linguistic features, and "hard-fusion" denotes the combination of all acoustic and linguistic features. In the Qwen2-Audio results, "MV" refers to the Majority Vote strategy across prompt types.}
    \begin{center}
        \begin{tabular}{|>{\centering\arraybackslash}p{0.45cm}|>{\centering\arraybackslash}p{0.5cm}|>{\centering\arraybackslash}p{1.1cm}|>{\centering\arraybackslash}p{1.2cm}|c|c|c|c|c|c|c|c|c|c|}
        \cline{2-14}
        \multicolumn{1}{c}{}& \multicolumn{6}{|c|}{\textbf{Qwen2-Audio}} & \multicolumn{7}{|c|}{\textbf{Baselines}} \\
        \cline{2-14}
             \multicolumn{1}{c|}{} &  Direct &	Contextual & Informative & CoT & CoT & MV & eGeMAPs & w2v & w2v & ling & w2v & ling & hard \\
            \multicolumn{1}{c|}{} &   &	 &  &  & Informative &  &  &  & +eGeMAPs &  & +ling & +eGeMAPs & fusion \\
        \hline
        UAR & 53.7 & 55.2 & 55.5 & 52.2 & 53.5 & \textbf{59} & 54.9 & 46.05 & \textbf{59.2} & 54.73 & 51.71 & 52.22 & 53.26 \\
        \hline
        \end{tabular}
    \end{center}
    \label{tab:results_tkd_test}
\end{table*}

\begin{figure*}[htbp]
\centering 
\includegraphics[width=1\textwidth]{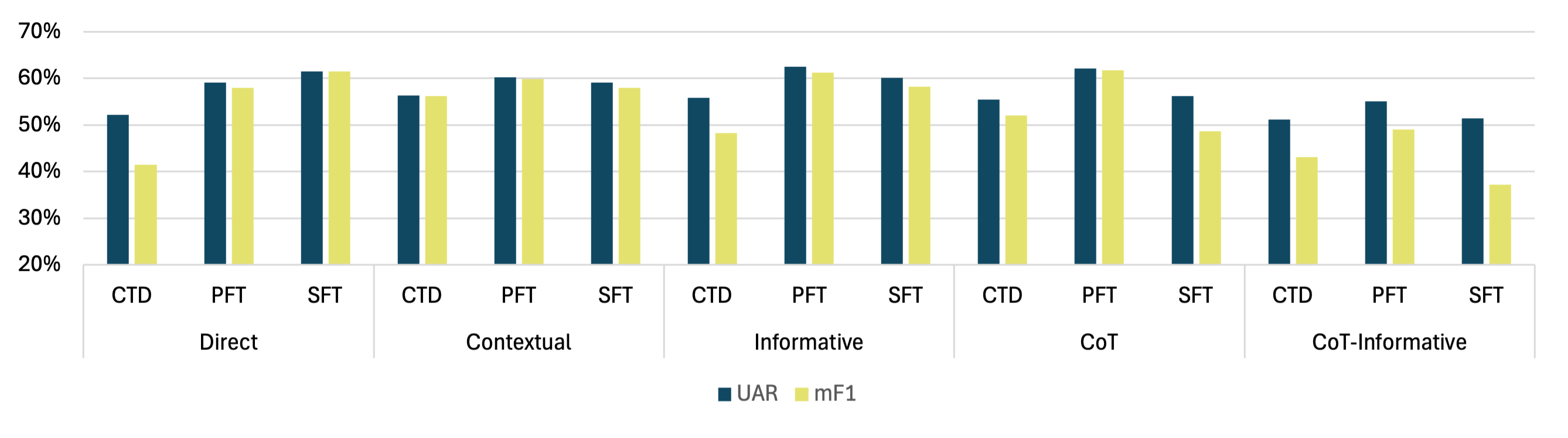} 
\caption{The Unweighted Average Recall (UAR) and macro F1 (mF1) scores for the PROCESS dataset across three tasks: Cookie Theft Description (CTD), Semantic Fluency Task (SFT), and Phonetic Fluency Task (PFT). CI detection from PFT achieved the highest performance, followed by SFT.}
\label{fig:results_process}
\end{figure*}

Fig. \ref{fig:results_process} presents the UAR and mF1 scores for the PROCESS dataset for the three tasks: Cookie Theft Description (CTD), Semantic Fluency Task (SFT), and Phonemic Fluency Task (PFT). Here, we also report the UAR of the best prompt for each prompt type in each task, highlighting the most effective prompt variations for different cognitive assessment tasks.

As shown in the figure, Contextual and Informative prompts outperform the Direct prompt in both CTD and PFT, reinforcing the finding from the TAUKADIAL dataset that adding contextual information improves classification in direct prompting strategies but does not provide an advantage in CoT prompting. Similarly, as observed in TAUKADIAL, CoT prompting did not improve performance, with UAR and macro-F1 decreasing in most tasks.

The best UAR performance for CTD was 56.3\%, which is comparable with the results from TAUKADIAL (~57\%) —both datasets share the picture description task. This indicates that picture description tasks yield consistent performance across datasets when using Qwen2-Audio AudioLLM.

On the other hand, the Fluency tasks (SFT and PFT) consistently outperformed CTD across almost all prompt strategies, suggesting that fluency-based tasks provide stronger cues for cognitive impairment classification using the AudioLLM model.

As discussed in Section \ref{sec:prompt}, we generated five reworded variants for each prompt type by prompting ChatGPT-4o to rewrite them while maintaining the same level of information and instructions. This aimed to assess the impact of rephrasing on the AudioLLM’s performance.

\begin{figure*}[htb]
\centering 
\includegraphics[width=1\textwidth]{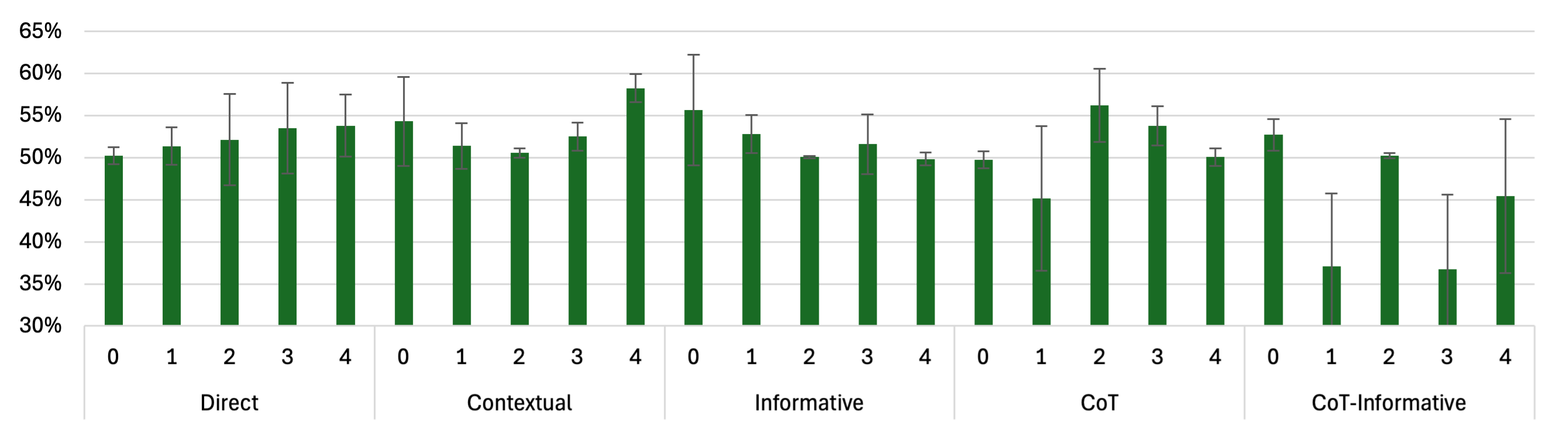} 
\caption{The impact of rewording prompts on Qwen2-Audio performance. Each prompt type has five variants (indexed 0–4). The bars represent the average UAR of each prompt across four datasets/tasks: TAUKADIAL, PROCESS-CTD, PROCESS-SFT, and PROCESS-PFT.}
\label{fig:results_prompts}
\end{figure*}

Fig. \ref{fig:results_prompts} illustrates the impact of prompt rewording on the AudioLLM’s performance. Each prompt type consists of five reworded variants (Prompt 0–4). Each variant was used in four separate experiments: TAUKADIAL, PROCESS-CTD, PROCESS-SFT, and PROCESS-PFT. The bar chart displays the average UAR for each prompt variant across all four experiments, providing insight into how rephrasing affects classification performance.

As shown in Fig. \ref{fig:results_prompts}, some prompt types are more sensitive to rewording than others. For example, in the CoT-Informative type, the second and fourth prompts exhibit significantly lower average UAR compared to the first prompt. In contrast, Direct prompts show relatively minor performance variations across different rewordings. Additionally, for most prompt types, a specific prompt tends to perform consistently better across both datasets and various tasks, indicating a degree of robustness in certain formulations.

\section{Conclusion}
In this paper, we proposed a zero-shot approach for detecting cognitive impairment (CI) from speech using Qwen2-Audio, a multimodal AudioLLM. We evaluated its performance on two datasets to assess its generalization and consistency across different tasks. Our findings suggest that the model’s classification is indeed based on analyzing speech for CI-related cues. Notably, the zero-shot classification achieved performance comparable to supervised baseline models.

Our results indicate that picture description tasks yield consistent performance across both datasets. Additionally, fluency-based tasks, such as semantic and phonetic fluency, proved more effective than picture description tasks. This may be because fluency tasks highlight acoustic markers of CI, such as long pauses, hesitations, and difficulty retrieving words, whereas picture description tasks rely more on linguistic features like sentence complexity, grammatical accuracy, and pragmatics.

Furthermore, while rewording prompts significantly impacts classification performance, we observed that the same prompt tends to yield consistent results across datasets and tasks. This suggests that certain prompts may generalize well for CI classification, reinforcing the potential of prompt engineering in zero-shot speech-based CI detection.

In future work, we plan to apply Qwen2-Audio to additional benchmark datasets such as ADReSSo \cite{Luz2020AlzheimersChallenge} and Pitt \cite{Becker1994TheDiagnosis} to further validate its effectiveness. Additionally, we will explore other AudioLLMs, including SALMONN \cite{Tang2024SALMONN:MODELS} and WAVLLM \cite{Hu2024WavLLM:Model}, to compare their performance in CI detection. Finally, we aim to enhance the effectiveness of AudioLLMs for CI detection, as these models are not originally trained for this task. One promising direction is to replace the general-purpose audio encoder with one fine-tuned to extract CI-relevant features, such as speech timing patterns and paralinguistic cues. Additionally, the entire model could be adapted using LoRA fine-tuning on diverse CI-related datasets, enabling task-specific learning while preserving the generalization capabilities of the underlying LLM.

\section*{Ethical Impact Statement}
This study did not involve any direct data collection or research with human subjects. All datasets used in this work were publicly available and previously collected in accordance with their respective ethical guidelines. As such, Institutional Review Board (IRB) approval was not required for this study.

Despite not involving human subject research, this work raises several ethical considerations related to the application of speech-based cognitive impairment (CI) detection models:
\begin{itemize}
    \item Bias and Generalizability: The performance of the AudioLLM model may be influenced by biases present in the training data of the underlying language and audio models, such as Qwen2-Audio. These biases could potentially lead to disparate performance across demographic groups, including age, gender, or language background. Moreover, one of the datasets used is English-only, and while the other is multilingual, the generalizability of the findings to languages or dialects not represented in these datasets may be limited.

    \item Misuse and Misinterpretation: Cognitive health is a sensitive domain, and automated CI detection models could be misused if deployed without appropriate oversight. There is a risk of these systems being interpreted as diagnostic tools rather than as preliminary screening aids. We stress that our system is not a replacement for clinical evaluation and should not be used for making medical decisions without human oversight.
\end{itemize}

To mitigate these risks, we advocate for cautious deployment of AudioLLMs in cognitive health applications and recommend that such tools be used in tandem with clinical expertise. We also encourage further research on the fairness, robustness, and interpretability of these models in diverse real-world contexts.

\bibliographystyle{IEEEtran}
\bibliography{references}

\begin{thebibliography}{10}
\providecommand{\url}[1]{#1}
\csname url@samestyle\endcsname
\providecommand{\newblock}{\relax}
\providecommand{\bibinfo}[2]{#2}
\providecommand{\BIBentrySTDinterwordspacing}{\spaceskip=0pt\relax}
\providecommand{\BIBentryALTinterwordstretchfactor}{4}
\providecommand{\BIBentryALTinterwordspacing}{\spaceskip=\fontdimen2\font plus
\BIBentryALTinterwordstretchfactor\fontdimen3\font minus
  \fontdimen4\font\relax}
\providecommand{\BIBforeignlanguage}[2]{{%
\expandafter\ifx\csname l@#1\endcsname\relax
\typeout{** WARNING: IEEEtran.bst: No hyphenation pattern has been}%
\typeout{** loaded for the language `#1'. Using the pattern for}%
\typeout{** the default language instead.}%
\else
\language=\csname l@#1\endcsname
\fi
#2}}
\providecommand{\BIBdecl}{\relax}
\BIBdecl

\bibitem{AssociationAlzheimers20232023Figures}
\BIBentryALTinterwordspacing
{Association Alzheimer’s}, ``{2023 Alzheimer's disease facts and figures},''
  Tech. Rep.~4, 4 2023. [Online]. Available:
  \url{https://onlinelibrary.wiley.com/doi/full/10.1002/alz.13016
  https://onlinelibrary.wiley.com/doi/abs/10.1002/alz.13016
  https://alz-journals.onlinelibrary.wiley.com/doi/10.1002/alz.13016}
\BIBentrySTDinterwordspacing

\bibitem{Nowrangi2016SubtleDatabase}
\BIBentryALTinterwordspacing
M.~A. Nowrangi, P.~B. Rosenberg, and J.~M.~S. Leoutsakos, ``{Subtle changes in
  daily functioning predict conversion from normal to mild cognitive impairment
  or dementia: an analysis of the NACC database},'' \emph{International
  psychogeriatrics}, vol.~28, no.~12, p. 2009, 12 2016. [Online]. Available:
  \url{https://pmc.ncbi.nlm.nih.gov/articles/PMC5628501/}
\BIBentrySTDinterwordspacing

\bibitem{Gamble2022CharacteristicsStudy}
\BIBentryALTinterwordspacing
L.~D. Gamble, F.~E. Matthews, I.~R. Jones, A.~E. Hillman, B.~Woods, C.~A.
  Macleod, A.~Martyr, R.~Collins, C.~Pentecost, J.~M. Rusted, and L.~Clare,
  ``{Characteristics of people living with undiagnosed dementia: findings from
  the CFAS Wales study},'' \emph{BMC Geriatrics}, vol.~22, no.~1, pp. 1--12, 12
  2022. [Online]. Available:
  \url{https://bmcgeriatr.biomedcentral.com/articles/10.1186/s12877-022-03086-4}
\BIBentrySTDinterwordspacing

\bibitem{Ding2024SpeechChallenges}
\BIBentryALTinterwordspacing
K.~Ding, M.~Chetty, {Azadeh}, N.~Hoshyar, {Tanusri Bhattacharya}, and B.~Klein,
  ``{Speech based detection of Alzheimer’s disease: a survey of AI
  techniques, datasets and challenges},'' \emph{Artificial Intelligence Review
  2024 57:12}, vol.~57, no.~12, pp. 1--43, 10 2024. [Online]. Available:
  \url{https://link.springer.com/article/10.1007/s10462-024-10961-6}
\BIBentrySTDinterwordspacing

\bibitem{DeLaFuenteGarcia2020ArtificialReview}
S.~De~La Fuente~Garcia, C.~W. Ritchie, and S.~Luz, ``{Artificial Intelligence,
  Speech, and Language Processing Approaches to Monitoring Alzheimer's Disease:
  A Systematic Review},'' \emph{Journal of Alzheimer's Disease}, vol.~78,
  no.~4, pp. 1547--1574, 2020.

\bibitem{Martinc2020TacklingDementia}
\BIBentryALTinterwordspacing
M.~Martinc and S.~Pollak, ``{Tackling the ADReSS challenge: a multimodal
  approach to the automated recognition of Alzheimer's dementia},'' in
  \emph{Interspeech}, 2020. [Online]. Available:
  \url{http://dx.doi.org/10.21437/Interspeech.2020-2202}
\BIBentrySTDinterwordspacing

\bibitem{Luz2020AlzheimersChallenge}
\BIBentryALTinterwordspacing
S.~Luz, F.~Haider, S.~De~La~Fuente, D.~Fromm, and B.~Macwhinney, ``{Alzheimer's
  Dementia Recognition through Spontaneous Speech: The ADReSS Challenge},'' in
  \emph{Interspeech}, 2020. [Online]. Available:
  \url{http://dx.doi.org/10.21437/Interspeech.2020-2571}
\BIBentrySTDinterwordspacing

\bibitem{Edwards2020MultiscaleSpeech}
\BIBentryALTinterwordspacing
E.~Edwards, C.~Dognin, B.~Bollepalli, and M.~Singh, ``{Multiscale System for
  Alzheimer's Dementia Recognition through Spontaneous Speech},'' in
  \emph{Interspeech}, 2020. [Online]. Available:
  \url{http://dx.doi.org/10.21437/Interspeech.2020-2781}
\BIBentrySTDinterwordspacing

\bibitem{Luz2024ConnectedEnglish}
\BIBentryALTinterwordspacing
S.~Luz, S.~De, L.~Fuente~Garcia, F.~Haider, D.~Fromm, B.~Macwhinney, A.~Lanzi,
  Y.-N. Chang, C.-J. Chou, and Y.-C. Liu, ``{Connected Speech-Based Cognitive
  Assessment in Chinese and English},'' in \emph{Interspeech}, 2024. [Online].
  Available: \url{http://luzs.gitlab.io/taukadial/}
\BIBentrySTDinterwordspacing

\bibitem{Ortiz-Perez2024CognitiveAnalysis}
D.~Ortiz-Perez, J.~Garcia-Rodriguez, and D.~Tom{\'{a}}s, ``{Cognitive Insights
  Across Languages: Enhancing Multimodal Interview Analysis},'' in
  \emph{Interspeech}.\hskip 1em plus 0.5em minus 0.4em\relax International
  Speech Communication Association, 9 2024, pp. 952--956.

\bibitem{AndreaPerez-Toro2024MultilingualChallenge}
\BIBentryALTinterwordspacing
P.~Andrea Perez-Toro, T.~Arias-Vergara, P.~Klumpp, T.~Weise, M.~Schuster,
  E.~Noth, J.~R. Orozco-Arroyave, and A.~Maier, ``{Multilingual Speech and
  Language Analysis for the Assessment of Mild Cognitive Impairment: Outcomes
  from the Taukadial Challenge},'' in \emph{Interspeech}, 2024. [Online].
  Available: \url{https://huggingface.co/facebook/}
\BIBentrySTDinterwordspacing

\bibitem{Becker1994TheDiagnosis}
\BIBentryALTinterwordspacing
J.~T. Becker, F.~Boiler, O.~L. Lopez, J.~Saxton, and K.~L. Mcgonigle, ``{The
  natural history of Alzheimer's disease. Description of study cohort and
  accuracy of diagnosis},'' \emph{Archives of neurology}, vol.~51, no.~6, pp.
  585--594, 1994. [Online]. Available:
  \url{https://pubmed.ncbi.nlm.nih.gov/8198470/}
\BIBentrySTDinterwordspacing

\bibitem{Tao2025EarlyChallenge}
\BIBentryALTinterwordspacing
F.~Tao, B.~Mirheidari, M.~Pahar, S.~Young, Y.~Xiao, H.~Elghazaly, F.~Peters,
  C.~Illingworth, D.~Braun, R.~O’Malley, S.~Bell, D.~Blackburn, F.~Haider,
  S.~Luz, and H.~Christensen, ``{Early Dementia Detection Using Multiple
  Spontaneous Speech Prompts: The PROCESS Challenge},'' in \emph{ICASSP 2025 -
  2025 IEEE International Conference on Acoustics, Speech and Signal Processing
  (ICASSP)}.\hskip 1em plus 0.5em minus 0.4em\relax IEEE, 4 2025, pp. 1--2.
  [Online]. Available: \url{https://ieeexplore.ieee.org/document/10889017/}
\BIBentrySTDinterwordspacing

\bibitem{Peng2024AMODELS}
J.~Peng, Y.~Wang, Y.~Xi, X.~Li, X.~Zhang, and K.~Yu, ``{A SURVEY ON SPEECH
  LARGE LANGUAGE MODELS},'' \emph{arXiv}, 2024.

\bibitem{Hu2024WavLLM:Model}
\BIBentryALTinterwordspacing
S.~Hu, L.~Zhou, S.~Liu, S.~Chen, L.~Meng, H.~Hao, J.~Pan, X.~Liu, J.~Li,
  S.~Sivasankaran, L.~Liu, and F.~Wei, ``{WavLLM: Towards Robust and Adaptive
  Speech Large Language Model},'' \emph{arXiv}, 3 2024. [Online]. Available:
  \url{http://arxiv.org/abs/2404.00656}
\BIBentrySTDinterwordspacing

\bibitem{Tang2024SALMONN:MODELS}
\BIBentryALTinterwordspacing
C.~Tang, W.~Yu, G.~Sun, X.~Chen, T.~Tan, W.~Li, L.~Lu, Z.~Ma, and C.~Zhang,
  ``{SALMONN: TOWARDS GENERIC HEARING ABILI-TIES FOR LARGE LANGUAGE MODELS},''
  in \emph{ICLR}, 2024. [Online]. Available:
  \url{https://github.com/bytedance/SALMONN.}
\BIBentrySTDinterwordspacing

\bibitem{Chu2023Qwen-Audio:Models}
\BIBentryALTinterwordspacing
Y.~Chu, J.~Xu, X.~Zhou, Q.~Yang, S.~Zhang, Z.~Yan, C.~Zhou, and J.~Zhou,
  ``{Qwen-Audio: Advancing Universal Audio Understanding via Unified
  Large-Scale Audio-Language Models},'' \emph{arXiv}, 2023. [Online].
  Available: \url{https://qwen-audio.github.io/}
\BIBentrySTDinterwordspacing

\bibitem{Chu2024Qwen2-AudioReport}
\BIBentryALTinterwordspacing
Y.~Chu, J.~Xu, Q.~Yang, H.~Wei Xipin Wei Zhifang Guo Yichong Leng Yuanjun Lv
  Jinzheng He Junyang Lin Chang~Zhou, and J.~Zhou, ``{Qwen2-Audio Technical
  Report},'' \emph{arXiv}, 2024. [Online]. Available:
  \url{https://github.com/QwenLM/Qwen2-Audio}
\BIBentrySTDinterwordspacing

\bibitem{Bai2023QwenReport}
\BIBentryALTinterwordspacing
J.~Bai, S.~Bai, Y.~Chu, Z.~Cui, K.~Dang, X.~Deng, Y.~Fan, W.~Ge, Y.~Han,
  F.~Huang, B.~Hui, L.~Ji, M.~Li, J.~Lin, R.~Lin, D.~Liu, G.~Liu, C.~Lu, K.~Lu,
  J.~Ma, R.~Men, X.~Ren, X.~Ren, C.~Tan, S.~Tan, J.~Tu, P.~Wang, S.~Wang,
  W.~Wang, S.~Wu, B.~Xu, J.~Xu, A.~Yang, H.~Yang, J.~Yang, S.~Yang, Y.~Yao,
  B.~Yu, H.~Yuan, Z.~Yuan, J.~Zhang, X.~Zhang, Y.~Zhang, Z.~Zhang, C.~Zhou,
  J.~Zhou, X.~Zhou, Z.~Q. Team, and A.~Group, ``{Qwen Technical Report},''
  \emph{arXiv}, 9 2023. [Online]. Available:
  \url{https://arxiv.org/abs/2309.16609v1}
\BIBentrySTDinterwordspacing

\bibitem{Cummins2020ARecognition}
\BIBentryALTinterwordspacing
N.~Cummins, Y.~Pan, Z.~Ren, J.~Fritsch, V.~S. Nallanthighal, H.~Christensen,
  D.~Blackburn, B.~W. Schuller, M.~M. Doss, H.~Strik, and A.~H{\"{a}}rm{\"{a}},
  ``{A Comparison of Acoustic and Linguistics Methodologies for Alzheimer's
  Dementia Recognition},'' in \emph{Interspeech}, 2020. [Online]. Available:
  \url{http://dx.doi.org/10.21437/Interspeech.2020-2635}
\BIBentrySTDinterwordspacing

\bibitem{Zafar2025Multi-ClassChallenge}
\BIBentryALTinterwordspacing
M.~A. Zafar, X.~Zhang, M.~Shahin, and B.~Ahmed, ``{Multi-Class Dementia
  Detection Using Acoustic Features - ICASSP-2025 PROCESS Challenge},''
  \emph{ICASSP 2025 - 2025 IEEE International Conference on Acoustics, Speech
  and Signal Processing (ICASSP)}, pp. 1--2, 4 2025. [Online]. Available:
  \url{https://ieeexplore.ieee.org/document/10889847/}
\BIBentrySTDinterwordspacing

\bibitem{Yuan2020DisfluenciesDisease}
\BIBentryALTinterwordspacing
J.~Yuan, Y.~Bian, X.~Cai, J.~Huang, Z.~Ye, and K.~Church, ``{Disfluencies and
  Fine-Tuning Pre-trained Language Models for Detection of Alzheimer's
  Disease},'' in \emph{Interspeech}, 2020. [Online]. Available:
  \url{http://dx.doi.org/10.21437/Interspeech.2020-2516}
\BIBentrySTDinterwordspacing

\bibitem{Baevski2020Wav2vecRepresentations}
\BIBentryALTinterwordspacing
A.~Baevski, H.~Zhou, A.~Mohamed, and M.~Auli, ``{wav2vec 2.0: A Framework for
  Self-Supervised Learning of Speech Representations},'' \emph{Advances in
  Neural Information Processing Systems}, vol. 2020-December, 2020. [Online].
  Available: \url{https://arxiv.org/abs/2006.11477v3}
\BIBentrySTDinterwordspacing

\bibitem{Hsu2021HuBERT:Units}
\BIBentryALTinterwordspacing
W.~N. Hsu, B.~Bolte, Y.~H.~H. Tsai, K.~Lakhotia, R.~Salakhutdinov, and
  A.~Mohamed, ``{HuBERT: Self-Supervised Speech Representation Learning by
  Masked Prediction of Hidden Units},'' \emph{IEEE/ACM Transactions on Audio
  Speech and Language Processing}, vol.~29, pp. 3451--3460, 6 2021. [Online].
  Available: \url{https://arxiv.org/abs/2106.07447v1}
\BIBentrySTDinterwordspacing

\bibitem{Radford2022RobustSupervision}
\BIBentryALTinterwordspacing
A.~Radford, J.~W. Kim, T.~Xu, G.~Brockman, C.~McLeavey, and I.~Sutskever,
  ``{Robust Speech Recognition via Large-Scale Weak Supervision},''
  \emph{Proceedings of Machine Learning Research}, vol. 202, pp.
  28\,492--28\,518, 12 2022. [Online]. Available:
  \url{https://arxiv.org/abs/2212.04356v1}
\BIBentrySTDinterwordspacing

\bibitem{Favaro2024LeveragingLanguages}
A.~Favaro, T.~Cao, N.~Dehak, and L.~Moro-Velazquez, ``{Leveraging Universal
  Speech Representations for Detecting and Assessing the Severity of Mild
  Cognitive Impairment Across Languages},'' in \emph{Interspeech}.\hskip 1em
  plus 0.5em minus 0.4em\relax International Speech Communication Association,
  9 2024, pp. 972--976.

\bibitem{Gosztolya2024CombiningChallenge}
G.~Gosztolya and L.~T{\'{o}}th, ``{Combining Acoustic Feature Sets for
  Detecting Mild Cognitive Impairment in the Interspeech'24 TAUKADIAL
  Challenge},'' in \emph{Interspeech}, 2024.

\bibitem{Kurtz2023EarlyCommands}
E.~Kurtz, Y.~Zhu, T.~Driesse, B.~Tran, J.~A. Batsis, R.~M. Roth, and X.~Liang,
  ``{Early Detection of Cognitive Decline Using Voice Assistant Commands},''
  \emph{ICASSP, IEEE International Conference on Acoustics, Speech and Signal
  Processing - Proceedings}, vol. 2023-June, 2023.

\bibitem{Bojanowski2017EnrichingInformation}
\BIBentryALTinterwordspacing
P.~Bojanowski, E.~Grave, A.~Joulin, and T.~Mikolov, ``{Enriching Word Vectors
  with Subword Information},'' \emph{Transactions of the Association for
  Computational Linguistics}, vol.~5, pp. 135--146, 12 2017. [Online].
  Available: \url{https://dx.doi.org/10.1162/tacl\_a\_00051}
\BIBentrySTDinterwordspacing

\bibitem{Pennington2014GloVe:Representation}
\BIBentryALTinterwordspacing
J.~Pennington, R.~Socher, and C.~D. Manning, ``{GloVe: Global Vectors for Word
  Representation},'' in \emph{EMNLP}, 2014, pp. 1532--1543. [Online].
  Available: \url{http://nlp.}
\BIBentrySTDinterwordspacing

\bibitem{Gupta2019BetterInformation}
\BIBentryALTinterwordspacing
P.~Gupta, M.~Pagliardini, and M.~Jaggi, ``{Better Word Embeddings by
  Disentangling Contextual n-Gram Information},'' \emph{NAACL 2019 - Conference
  of the North American Chapter of the Association for Computational
  Linguistics: Human Language Technologies}, vol.~1, pp. 933--939, 2019.
  [Online]. Available:
  \url{https://infoscience.epfl.ch/handle/20.500.14299/159030}
\BIBentrySTDinterwordspacing

\bibitem{Barrera-Altuna2024TheApproach}
B.~Barrera-Altuna, D.~Lee, Z.~Zarnaz, J.~Han, and S.~Kim, ``{The Interspeech
  2024 TAUKADIAL Challenge: Multilingual Mild Cognitive Impairment Detection
  with Multimodal Approach},'' in \emph{Interspeech}.\hskip 1em plus 0.5em
  minus 0.4em\relax International Speech Communication Association, 9 2024, pp.
  967--971.

\bibitem{Hoang2024TranslingualSpeech}
B.~Hoang, Y.~Pang, H.~Dodge, and J.~Zhou, ``{Translingual Language Markers for
  Cognitive Assessment from Spontaneous Speech},'' in \emph{Interspeech}.\hskip
  1em plus 0.5em minus 0.4em\relax International Speech Communication
  Association, 9 2024, pp. 977--981.

\bibitem{Balagopalan2020ToDetection}
\BIBentryALTinterwordspacing
A.~Balagopalan, B.~Eyre, F.~Rudzicz, J.~Novikova, and K.~Li, ``{To BERT or Not
  To BERT: Comparing Speech and Language-based Approaches for Alzheimer's
  Disease Detection},'' in \emph{Interspeech}, 2020. [Online]. Available:
  \url{http://dx.doi.org/10.21437/Interspeech.2020-2557}
\BIBentrySTDinterwordspacing

\bibitem{Rohanian2020Multi-modalSpeech}
\BIBentryALTinterwordspacing
M.~Rohanian, J.~Hough, and M.~Purver, ``{Multi-modal fusion with gating using
  audio, lexical and disfluency features for Alzheimer's Dementia recognition
  from spontaneous speech},'' in \emph{Interspeech}, 2020. [Online]. Available:
  \url{http://dx.doi.org/10.21437/Interspeech.2020-2721}
\BIBentrySTDinterwordspacing

\bibitem{Duan2024Pre-trainedDetection}
\BIBentryALTinterwordspacing
J.~Duan, F.~Wei, H.~Li, and J.~Liu, ``{Pre-trained Feature Fusion and Matching
  for Mild Cognitive Impairment Detection},'' 2024. [Online]. Available:
  \url{https://github.com/facebookresearch/}
\BIBentrySTDinterwordspacing

\bibitem{Goodglass1983BostonBooklet}
H.~Goodglass and E.~Kaplan, \emph{{Boston diagnostic aphasia examination
  booklet}}.\hskip 1em plus 0.5em minus 0.4em\relax Lea {\textbackslash}{\&}
  Febiger, 1983.

\bibitem{Vaswani2017AttentionNeed}
\BIBentryALTinterwordspacing
A.~Vaswani, G.~Brain, N.~Shazeer, N.~Parmar, J.~Uszkoreit, L.~Jones, A.~N.
  Gomez, L.~Kaiser, and I.~Polosukhin, ``{Attention Is All You Need},'' in
  \emph{NIPS}, 2017. [Online]. Available:
  \url{https://dl.acm.org/doi/10.5555/3295222.3295349}
\BIBentrySTDinterwordspacing

\bibitem{Bahdanau2014NeuralTranslate}
\BIBentryALTinterwordspacing
D.~Bahdanau, K.~H. Cho, and Y.~Bengio, ``{Neural Machine Translation by Jointly
  Learning to Align and Translate},'' \emph{3rd International Conference on
  Learning Representations, ICLR 2015 - Conference Track Proceedings}, 9 2014.
  [Online]. Available: \url{https://arxiv.org/abs/1409.0473v7}
\BIBentrySTDinterwordspacing

\bibitem{Devlin2018BERT:Understanding}
\BIBentryALTinterwordspacing
J.~Devlin, M.~W. Chang, K.~Lee, and K.~Toutanova, ``{BERT: Pre-training of Deep
  Bidirectional Transformers for Language Understanding},'' \emph{NAACL HLT
  2019 - 2019 Conference of the North American Chapter of the Association for
  Computational Linguistics: Human Language Technologies - Proceedings of the
  Conference}, vol.~1, pp. 4171--4186, 10 2018. [Online]. Available:
  \url{https://arxiv.org/abs/1810.04805v2}
\BIBentrySTDinterwordspacing

\bibitem{Openai2018ImprovingPre-Training}
\BIBentryALTinterwordspacing
A.~R. Openai, K.~N. Openai, T.~S. Openai, and I.~S. Openai, ``{Improving
  Language Understanding by Generative Pre-Training},'' 2018. [Online].
  Available: \url{https://gluebenchmark.com/leaderboard}
\BIBentrySTDinterwordspacing

\bibitem{Thirunavukarasu2023LargeMedicine}
\BIBentryALTinterwordspacing
A.~J. Thirunavukarasu, D.~S.~J. Ting, K.~Elangovan, L.~Gutierrez, T.~F. Tan,
  and D.~S.~W. Ting, ``{Large language models in medicine},'' \emph{Nature
  Medicine 2023 29:8}, vol.~29, no.~8, pp. 1930--1940, 7 2023. [Online].
  Available: \url{https://www.nature.com/articles/s41591-023-02448-8}
\BIBentrySTDinterwordspacing

\bibitem{Savage2024DiagnosticMedicine}
\BIBentryALTinterwordspacing
T.~Savage, A.~Nayak, R.~Gallo, E.~Rangan, and J.~H. Chen, ``{Diagnostic
  reasoning prompts reveal the potential for large language model
  interpretability in medicine},'' \emph{npj Digital Medicine 2024 7:1},
  vol.~7, no.~1, pp. 1--7, 1 2024. [Online]. Available:
  \url{https://www.nature.com/articles/s41746-024-01010-1}
\BIBentrySTDinterwordspacing

\bibitem{Chen2021WavLM:Processing}
\BIBentryALTinterwordspacing
S.~Chen, C.~Wang, Z.~Chen, Y.~Wu, S.~Liu, Z.~Chen, J.~Li, N.~Kanda,
  T.~Yoshioka, X.~Xiao, J.~Wu, L.~Zhou, S.~Ren, Y.~Qian, Y.~Qian, J.~Wu,
  M.~Zeng, X.~Yu, and F.~Wei, ``{WavLM: Large-Scale Self-Supervised
  Pre-Training for Full Stack Speech Processing},'' \emph{IEEE Journal on
  Selected Topics in Signal Processing}, vol.~16, no.~6, pp. 1505--1518, 10
  2021. [Online]. Available: \url{http://arxiv.org/abs/2110.13900
  http://dx.doi.org/10.1109/JSTSP.2022.3188113}
\BIBentrySTDinterwordspacing

\bibitem{Hu2021LORA:MODELS}
\BIBentryALTinterwordspacing
E.~Hu, Y.~Shen, P.~Wallis, Z.~Allen-Zhu, Y.~Li, S.~Wang, L.~Wang, and W.~Chen,
  ``{LORA: LOW-RANK ADAPTATION OF LARGE LAN-GUAGE MODELS},'' \emph{arXiv},
  2021. [Online]. Available: \url{https://github.com/microsoft/LoRA.}
\BIBentrySTDinterwordspacing

\bibitem{Chen2022BEATs:Tokenizers}
\BIBentryALTinterwordspacing
S.~Chen, Y.~Wu, C.~Wang, S.~Liu, D.~Tompkins, Z.~Chen, W.~Che, X.~Yu, and
  F.~Wei, ``{BEATs: Audio Pre-Training with Acoustic Tokenizers},''
  \emph{Proceedings of Machine Learning Research}, vol. 202, pp. 4672--4712, 12
  2022. [Online]. Available: \url{https://arxiv.org/abs/2212.09058v1}
\BIBentrySTDinterwordspacing

\bibitem{Zheng2023JudgingArena}
\BIBentryALTinterwordspacing
L.~Zheng, W.-L. Chiang, Y.~Sheng, S.~Zhuang, Z.~Wu, Y.~Zhuang, Z.~Lin, Z.~Li,
  D.~Li, E.~P. Xing, H.~Zhang, J.~E. Gonzalez, and I.~Stoica, ``{Judging
  LLM-as-a-Judge with MT-Bench and Chatbot Arena},'' in \emph{Neural
  Information Processing Systems}, 6 2023. [Online]. Available:
  \url{https://arxiv.org/abs/2306.05685v4}
\BIBentrySTDinterwordspacing

\bibitem{Wang2024AudioBench:Models}
\BIBentryALTinterwordspacing
B.~Wang, X.~Zou, G.~Lin, S.~Sun, Z.~Liu, W.~Zhang, Z.~Liu, A.~Aw, and N.~F.
  Chen, ``{AudioBench: A Universal Benchmark for Audio Large Language
  Models},'' \emph{arXiv}, 6 2024. [Online]. Available:
  \url{http://arxiv.org/abs/2406.16020}
\BIBentrySTDinterwordspacing

\bibitem{Chen2024VoiceBench:Assistants}
\BIBentryALTinterwordspacing
Y.~Chen, X.~Yue, C.~Zhang, X.~Gao, R.~T. Tan, and H.~Li, ``{VoiceBench:
  Benchmarking LLM-Based Voice Assistants},'' \emph{arXiv}, 2024. [Online].
  Available: \url{https://github.com/}
\BIBentrySTDinterwordspacing

\bibitem{Qwen2024Qwen2.5Report}
\BIBentryALTinterwordspacing
{Qwen}, {:}, A.~Yang, B.~Yang, B.~Zhang, B.~Hui, B.~Zheng, B.~Yu, C.~Li,
  D.~Liu, F.~Huang, H.~Wei, H.~Lin, J.~Yang, J.~Tu, J.~Zhang, J.~Yang, J.~Yang,
  J.~Zhou, J.~Lin, K.~Dang, K.~Lu, K.~Bao, K.~Yang, L.~Yu, M.~Li, M.~Xue,
  P.~Zhang, Q.~Zhu, R.~Men, R.~Lin, T.~Li, T.~Tang, T.~Xia, X.~Ren, X.~Ren,
  Y.~Fan, Y.~Su, Y.~Zhang, Y.~Wan, Y.~Liu, Z.~Cui, Z.~Zhang, and Z.~Qiu,
  ``{Qwen2.5 Technical Report},'' \emph{arXiv}, 12 2024. [Online]. Available:
  \url{https://arxiv.org/abs/2412.15115v2}
\BIBentrySTDinterwordspacing

\end{thebibliography}

\end{document}